\documentclass{desyproc}
\begin{document}
\title{\kern-24pt\vbox to0pt{\vss\hbox to0pt{\vtop to0pt{\vss\kern-12pt
  \hbox to\textwidth{\normalsize\rm\hfill MITP/14-004}\kern-10pt
  \hbox to\textwidth{\normalsize\rm\hfill February 2014}}\hss}}%
  Helicity Amplitudes and\\ Angular Decay
  Distributions\footnote{Lecture given at the Helmholtz International Summer
School ``Physics of Heavy Quarks and Hadrons'', Dubna, Russia, July 15--28,
2013 (to be published in the Proceedings)}}

\author{{\slshape J.G.~K\"orner$^1$}\\[1ex]
$^1$PRISMA Cluster of Excellence, Institut f\"ur Physik,
  Johannes-Gutenberg-Universit\"at,\\[.2cm]
  Staudinger Weg 7, 55099 Mainz, Germany }

\contribID{xy}

\desyproc{DESY-PROC-2013-03}
\acronym{HQ2013}

\maketitle

\begin{abstract}
I discuss how to obtain angular decay distributions for sequential cascade
decays using helicity methods. The angular decay distributions follow from a
reasonably simple master formula involving bilinear forms of helicity
amplitudes and Wigner's $d$ functions. I discuss in some detail the issue of
gauge invariance for off-shell gauge bosons. As a technical exercise I
calculate the linear relation between the helicity amplitudes and the
invariant amplitudes of semileptonic and rare baryon decays. I discuss two
explicit examples of angular decay distributions for (i) the decay
$t\to b+W^+(\to \ell^+\nu_\ell)$ (which leads to the notion of the helicity
fractions of the $W^+$), and (ii) the sequential decay
$\Lambda_b\to\Lambda(\to p\pi^-)+J/\psi(\to \ell^+\ell^-)$. 
\end{abstract}

\section{Introductory remarks}
In these lectures I want to discuss some examples of sequential cascade decays
and their corresponding angular decay distributions. The angular decay
distributions follow from a reasonably simple master formula involving
bilinear forms of helicity amplitudes and Wigner's $d$ functions.
Some sample cascade decay processes are
\begin{itemize}
\item Polarized top quark decay~\cite{Fischer:1998gsa,Fischer:2001gp}\quad
  $t(\uparrow)\to b+W^+(\to\ell^+\nu_\ell)$
\item Rare $\Lambda_b(\uparrow)$ decays~\cite{Gutsche:2013pp}\quad
  $\Lambda_b(\uparrow)\to \Lambda_s(\to p\pi^-)+j_{\rm eff}(\to \ell^+\ell^-)$
\item Higgs decay to gauge bosons~\cite{berge14}\quad
  $H\to W^+(\to \ell^+\nu_\ell)+W^{-\ast}(\to\ell^-\bar\nu_\ell)$\\
  \hspace*{5.2cm}$H\to Z(\to\ell^+\ell^-)+Z^\ast(\to\ell^+\ell^-)$
\item Rare $B$ decays~\cite{Faessler:2002ut}\quad
  $B\to D+j_{\rm eff}(\to\ell^+\ell^-)$\\\hspace*{3.0cm}
  $B\to D^\ast(\to D\pi)+j_{\rm eff}(\to\ell^+\ell^-)$
\item Semileptonic $\Lambda_b$ decays~\cite{Korner:1991ph}\quad
  $\Lambda_b\to\Lambda_c(\to\Lambda\pi^+)+W^{-\ast}(\to\ell^-\bar\nu_\ell)$ 
\item Semileptonic $B$ decays~\cite{Korner:1987kd,Korner:1989ve,Korner:1989qb}
  \quad $B\to D+W^\ast(\to\ell\nu)$\\\hspace*{5.1cm}
  $B\to D^{\ast}(\to D\pi)+W^\ast(\to\ell\nu)$
\item Nonleptonic $\Lambda_b$ decays~\cite{Gutsche:2013oea}\quad
  $\Lambda_b\to\Lambda(\to p\pi^-)+J/\psi(\to\ell^+\ell^-)$
\item Semileptonic hyperon decays ($\ell^-=e^-,\mu^-$)~\cite{Kadeer:2005aq}
  \quad
  $\Xi^0(\uparrow)\to\Sigma^+(\to p+\pi^0)+W^{-\ast}(\to\ell^-\bar\nu_\ell)$ 
\end{itemize}
In our treatment of these cascade decay processes we have accounted for 
lepton mass effects whenever this is warranted for by the decay kinematics.

The generic form of most of the above cascade decays is
$H_1\to H_2(\to H_3+H_4)+W,W^\ast$, $j_{\rm eff}(\to\ell+\bar\ell)$ where the
$H_i$ can be mesons, baryons or quarks, and the $W$ and $W^\ast$ denote either
on-shell or off-shell charged $W$'s. For neutral current transitions,
$j_{\rm eff}$ denotes an effective four-vector and/or four-axial vector
current relevant for the desription of rare decays. The interest in deriving
angular decay distributions via helicity methods is two-fold. First it
facilitates the theoretical analysis of a decay distribution in terms of e.g.\
parity or CP violating contributions. Second it allows one to generate
experimental decay distributions via a suitable Monte Carlo program 
(see e.g.\ Ref.~\cite{Kadeer:2005aq}).

Take as an example the semileptonic hyperon decay 
$\Xi^0(\uparrow)\to\Sigma^+(\to p+\pi^0)+\ell^-+\bar\nu_\ell$
($\ell^-=e^-,\mu^-$). The decay process is described by three polar angles
$\theta$, $\theta_B$ and $\theta_P$ (as e.g.\ in Fig.~\ref{three-fold}) and
two azimuthal angles $\phi_B$  and $\phi_\ell$ which describe the relative
azimuthal orientation of the two planes that characterize the cascade decay
process.

As we shall learn in this lecture, the angular decay distribution can be
derived from the master formula~\cite{Kadeer:2005aq}
\begin{eqnarray}\label{five1}
W(\theta,\theta_P,\theta_B,\phi_B,\phi_\ell)&\propto&
  \kern-36pt\sum_{\lambda_\ell,\lambda_W,\lambda'_W,J,J',\lambda_2,
  \lambda_2',\lambda_3}\kern-36pt
  (-1)^{J+J'}|h^{V-A}_{\lambda_l\lambda_{\nu=\pm 1/2}}|^2
  e^{i(\lambda_W-\lambda'_W)\phi_\ell}\\&&\times\
  \rho_{\lambda_2-\lambda_W,\lambda'_2-\lambda'_W}(\theta_P)
  d^J_{\lambda_W,\lambda_\ell-\lambda_\nu}(\theta)
  d^{J'}_{\lambda'_W,\lambda_\ell-\lambda_\nu}(\theta)
  H_{\lambda_2\lambda_W}H^*_{\lambda'_2\lambda'_W}\nonumber\\&&\times\
  e^{i(\lambda_2-\lambda'_2)\phi_B}d^{1/2}_{\lambda_2\lambda_3}(\theta_B)
  d^{1/2}_{\lambda_2'\lambda_3}(\theta_B)|h^B_{\lambda_3 0}|^2\nonumber
\end{eqnarray}
where
\begin{eqnarray*}
h^{V-A}_{\lambda_\ell\lambda_{\nu=\pm 1/2}} &:&\qquad
\mbox{helicity amplitudes for the transition $W^\ast\to\ell+\nu_\ell$}\,:
  \\&&\qquad
  \mbox{$\lambda_{\bar\nu}=1/2$ for $(\ell^-\bar\nu_\ell)$;
  $\lambda_\nu=-1/2$ for $(\ell^+\nu_\ell)$}\\
\rho_{\lambda_1\lambda'_1}&:&\qquad
  \mbox{density matrix for the polarized parent baryon $B_1$}\\
H_{\lambda_2\lambda_W}&:&\qquad
  \mbox{helicity amplitudes for the transition $B_1\to B_2+W^\ast$}\\
h^B_{\lambda_3 0}&:&\qquad
  \mbox{helicity amplitudes for the transition $B_2\to B_3+\pi$}\\
d^J_{mm'}&:&\qquad
  \mbox{Wigner's $d$ functions}
\end{eqnarray*}

The $\lambda_{i},\,\lambda_{\ell},\,\lambda_{W},\,\ldots$ are helicity labels 
of the baryons, leptons and the $W^\ast$ that participate in the process. 
They take the values
\begin{eqnarray*}
\lambda_1,\,\lambda_2,\,\lambda_3,\,\lambda_\ell&=&\pm1/2\\
\lambda_W&=&1,\,0,-1\,\,(J=1);\quad t \,\,(J=0)\\
\lambda_{\bar\nu}&=& +1/2;\,\lambda_\nu=-1/2
\end{eqnarray*}

We shall see in these lectures that the off-shell gauge boson $W$ has a
spin-$1$ and a spin-$0$ component. Thus we have to sum over $J=0,\,1$. The
phase factor $(-1)^{J+J'}=\pm1$ is associated with the  Minkowski metric of
our world. The angular decay distribution~(\ref{five1}) covers both final
lepton states $(\ell^-\bar\nu_\ell)$ and $(\ell^+\nu_\ell)$ which are
distinguished through the labelling $\lambda_\nu=\pm1/2$
($\lambda_{\bar\nu}=+1/2$, $\lambda_\nu=-1/2$). This covers the charge
conjugated process or also the semileptonic decay
$\Sigma^+\to\Lambda+e^+\nu_e$.

The master formula~(\ref{five1}) is quite general. After appropiate angular 
integrations over $\theta_B$ and $\phi_B$ the master formula also applies to
the three-fold angular decay distribution of polarized top decay 
$t(\uparrow)\to b+\ell^+\nu_\ell$, etc., etc.. The summation over helicities
can be quite elaborate if done by hand. However, the summation can be done by
computer. A FORM package doing the summation automatically is available from
M.A.~Ivanov.

\subsection{Polarization of the lepton}
In the master formula~(\ref{five1}) I have summed over the helicities of the
lepton. To obtain the polarization of the lepton leave the lepton helicity
unsummed, i.e.\
\begin{equation}
\sum_{\lambda_\ell,\,\ldots}\qquad \to \qquad \sum_{\ldots}\nonumber
\end{equation}
For example, the longitudinal polarization of the charged lepton is then given
by
\begin{equation}\label{leppol}
P^z(\ell)=\frac{W_{\lambda_\ell=+1/2}-W_{\lambda_\ell=-1/2}}
  {W_{\lambda_\ell=+1/2}+W_{\lambda_\ell=-1/2}}
\end{equation}
In the same vein the transverse polarization components $P^x(\ell)$ and
$P^y(\ell)$ can be obtained from the nondiagonal elements of the $W^\ast$
density matrix. Note that the longitudinal polarization of the lepton
in Eq.~(\ref{leppol}) refers to the lepton-neutrino cm system, and {\em not}
to the $\Xi^0$ rest system.   

\section{Gauge boson off-shell effects}
\subsection{Off-shell effects and scalar degrees of freedom}
When the gauge boson is off its mass shell $q^2\neq m_{W,Z}^2$ one has to take
into account the scalar degree of freedom of the gauge boson. Take the unitary
gauge and write out the numerator of the $W$ gauge boson propagator as
\begin{equation}
H_{\mu\nu}L^{\mu\nu}=H_{\mu\nu}\,g^{\mu\mu'}g^{\nu\nu'}\,L_{\mu'\nu'}
  \quad\longrightarrow\quad H_{\mu\nu}\,
  \Big(g^{\mu\mu'}-\frac{q^{\mu}q^{\mu'}}{m_W^2}\Big)
  \Big(g^{\nu\nu'}-\frac{q^{\nu}q^{\nu'}}{m_W^2}\Big)\,L_{\mu'\nu'}.
\nonumber
\end{equation}
The term $q^{\mu}q^{\mu'}/m_W^2$ is usually dropped in low energy applications
such as $\mu$-decay and in semileptonic decays in the charm and bottom 
sector. Split the
propagator numerator into a spin-$1$ and a spin-$0$ piece
\begin{equation}
\Big(
\underbrace{-g^{\mu\mu'}+\frac{q^{\mu}q^{\mu'}}{q^2}}_{\rm spin\,1}
  -\underbrace{\frac{q^{\mu}q^{\mu'}}{q^2}
  (1-\frac{q^2}{m_W^2})}_{\rm spin\,0}\Big)\Big(
\underbrace{-g^{\nu\nu'}+\frac{q^{\nu}q^{\nu'}}{q^2}}_{\rm spin\,1}
  -\underbrace{\frac{q^{\nu}q^{\nu'}}{q^2}
  (1-\frac{q^2}{m_W^2})}_{\rm spin\,0}\Big).\nonumber
\end{equation}
There are three contributions
$i)\,\,{\rm spin}\,1 \otimes {\rm spin}\,1$,\,
$ii)\,-\,(\,{\rm spin}\,1 \otimes {\rm spin}\,0 + 
{\rm spin}\,0 \otimes {\rm spin}\,1\,)$ and
$iii)\,\,{\rm spin}\,0 \otimes {\rm spin}\,0$.

Note the minus sign in case ii) which results from the Minkowski metric. This
extra minus sign can be readily incorporated into the master formulas for
angular decay distributions by introducing the factor $(-1)^{J+J'}$ and
summing over $J,J'=0,\,1$. The scalar contributions are $O(m_\ell^2)$ since 
$q^\mu L_{\mu\nu}\sim O(m_\ell)$. Note, however, that $q^2$ can be small since
the range of off-shellness is
\begin{equation}
(m_{\ell\,1}+m_{\ell\,2})^{2}\le q^2\le (M_1-M_2)^2\nonumber
\end{equation}
for the decay $H_1(M_1)\to H_2(M_2)+\ell_1(m_{\ell_1})+\bar\ell_2(m_{\ell_2})$.

\subsection{The issue of gauge invariance}
Consider the gauge boson propagator in the general $R_\xi$ gauge and rewrite
it into a convenient form. For definiteness we consider the decay
$t\to b+W^+$ where we shall also consider gauge boson off-shell effects which
allows one to calculate finite width effects as will be done in Sec.~2.3.
\begin{eqnarray}\label{rxi}
D^{\mu\nu}&=&\frac{i}{q^2-m_W^2}\left(-g^{\mu\nu}
  +\frac{q^{\mu}q^{\nu}(1-\xi_W)}{q^2-\xi_W m_W^2}\right)\\
  &=&\frac{i}{q^2-m_W^2}\left(-g^{\mu\nu}+\frac{q^{\mu} q^{\nu}}{m_W^2}
  -\frac{q^{\mu}q^{\nu}}{m_W^2}
  +\frac{q^{\mu}q^{\nu}(1-\xi_W)}{q^2-\xi_W m_W^2}\right)\nonumber 
\end{eqnarray}
resulting in 
\begin{equation}\label{rxi12}
D^{\mu\nu}=\frac{i}{q^2-m_W^2}
\left(-g^{\mu\nu}+\frac{q^{\mu}q^{\nu}}{m_W^2}\right)
-i\,\frac{q^{\mu}q^{\nu}}{m_W^2}\,\frac1{q^2-\xi_W m_W^2}.
\end{equation}
The first term in Eq.~(\ref{rxi12}) is referred to as the unitary propagator.
The second gauge-dependent term in Eq.~(\ref{rxi12}) can be seen to exactly
cancel the contribution of the charged Goldstone $\phi^+$ exchange if fermion
lines are attached to the gauge boson and the charged Goldstone boson
contribution. One uses the Dirac equation to convert the $q^{\mu}$ and
$q^{\nu}$ contributions in the second term of Eq.~(\ref{rxi12}) to fermion 
masses. In our case one would have
\begin{eqnarray*}
q^\nu\bar u_f\gamma_\nu(1-\gamma_5)v_{\bar f}
  &=&m_f\bar u_f(1-\gamma_5)v_{\bar f}
  +m_{\bar f}\bar u_f(1+\gamma_5)v_{\bar f},\\[7pt]
q^\mu\bar u_b\gamma_\mu(1-\gamma_5)u_t
  &=&m_t\bar u_b(1+\gamma_5)u_t
  -m_b\bar u_b(1-\gamma_5)u_t.
\end{eqnarray*}
One can then see that the second term in Eq.~(\ref{rxi12}) is exactly
cancelled by the corresponding $\phi^+$-exchange contribution (with the same
fermion pair $(f\bar f)$ attached). This exercise shows that it does not make
sense to talk of an external off-shell gauge boson in isolation. One must
include the coupling to a final state fermion pair if one wants to obtain a
gauge invariant result. 

\subsection{Off-shell effects in the decay $t\to b + W^{+}$}
In the zero width approximation and using the unitary gauge the differential
rate for $t\to b+W^+$ is given by
\begin{equation}\label{topdecay}
\frac{d\Gamma}{dq^2}\sim H_{\mu\nu}\,
  \Big(g^{\mu\mu'}-\frac{q^{\mu}q^{\mu'}}{m_W^2}\Big)
  \Big(g^{\nu\nu'}-\frac{q^{\nu}q^{\nu'}}{m_W^2}\Big)\,L_{\mu'\nu'}\,
  \delta(q^2-m_W^2).\nonumber
\end{equation}
On shell one has $q^2=m_W^2$, and it makes no difference whether one uses the
Landau gauge ($\xi=0$) with $(g^{\mu\nu}-q^{\mu}q^{\nu}/q^2)$ or the unitary
gauge ($\xi=\infty$) with $(g^{\mu\nu}-q^{\mu}q^{\nu}/m_W^2)$. Since we want
to account for off-shell effects the use of the unitary gauge is mandatory as
explained in Sec.~2.2. Finite width effects can be accounted for by smearing
the zero-width formula with the replacement
\begin{equation}
   \delta(q^2 - m_W^2)\quad\longrightarrow\quad 
    \frac{m_W\Gamma_W}{\pi}\frac1{(q^2-m_W^2)^2+m_W^2\Gamma_W^2}.\nonumber
\end{equation}
One then integrates in the limits
\begin{equation}
m_\ell^2\le q^2\le(m_t-m_b)^2.\nonumber
\end{equation}
Numerically the finite width corrections amount to $-1.55\%$\, in
$\Gamma_{t\to b+W^+}$~\cite{Jezabek:1993wk,Do:2002ky}. Curiously, the negative
finite width corrections are almost completely cancelled by the positive first
order electroweak corrections~\cite{Do:2002ky}.

\subsection{Scalar contributions in some sample decay processes}
Scalar contributions are of $O(m_\ell^2)$. They are therefore important for
decay processes where the lepton mass is comparable to the scale of the decay
process. For semileptonic and rare processes the characteristic scale would be
given by the mass difference $M_1-M_2$. A more symmetric scale is used in the
PDG tables, namely the largest momentum of any of the decay products in the
rest frame of the decaying particle. Sample decay processes and their scalar
contributions are
\begin{itemize}
\item Decays involving the $\tau$
\begin{eqnarray*}
B\to D+\tau\nu_\tau\, &:&\qquad \Gamma_S/\Gamma\,\approx\,58\,\%
  \quad\cite{Korner:1989ve,Korner:1989qb} \\
B\to D^\ast+\tau\nu_\tau\,&:&\qquad \Gamma_S/\Gamma\,\approx\,7\,\% 
  \quad\cite{Korner:1989ve,Korner:1989qb}\\
B\to\pi+\tau\nu_\tau\, &:&\qquad \Gamma_S/\Gamma\,\approx\,(30\div 50)\,\% 
  \quad\cite{Dominguez:1990mi}\\ 
H\to W^+W^{-\ast}(\to\tau^-\nu_\tau)&:&\qquad \Gamma_S/\Gamma\,=\,0.73\,\%
  \quad\cite{berge14}\\
H\to ZZ^\ast(\to\tau^+\tau^-)&:&\qquad \Gamma_S/\Gamma\,=\,1.19\,\%
  \quad\cite{berge14}
\end{eqnarray*}
The decays $B\to D^{(\ast)}+\tau\nu_\tau$ and $B\to\pi+\tau\nu_\tau$ have been
widely discussed in the literature because the scalar contribution can be
augmented by charged Higgs exchange~\cite{Nierste:2008qe,Fajfer:2012vx}.
\item Hadronic semi-inclusive decays $H\to ZZ^\ast(\to b\bar b)$
\begin{equation}
H\to ZZ^\ast(\to b\bar b)\quad:\quad\qquad \Gamma_S/\Gamma\,=\,7.9\,\%
  \quad\cite{Groote:2013hc}\nonumber
\end{equation}
\end{itemize}
Since the ratio $m_e/(m_n-m_p)=0.395$ is not small it comes of no surprise
that there is a sizeable scalar contribution to the neutron $\beta$ decay
$n\to p+e^-\bar\nu_e$. In fact one finds $\Gamma_S/\Gamma\,=\,19\,\%$ for
neutron $\beta$ decay.

\subsection{Scalar contribution to the FB asymmetry $A_{FB}$
  of the lepton pair}
An interesting observation concerns the scalar contribution to the
Forward-Backward (FB) asymmetry of the lepton pair in the cm frame of the
lepton pair or, put differently, in the $W^\ast$ rest frame where its momentum
is $(\sqrt{q^2},\vec 0\,)$. The notation $\vec 0$ is rather symbolic and
stands for the momentum direction of the $W$ before it is boosted to its 
rest frame. The
observation is that there are parity-conserving contributions to the
FB asymmetry arising from scalar-vector interference effects.
 Consider the FB asymmetry 
\begin{equation}\label{fb}
A_{FB}=\frac{\Gamma_F-\Gamma_B}{\Gamma_F+\Gamma_B}
\end{equation}
\begin{figure}[!htb]
\begin{center}
\epsfig{figure=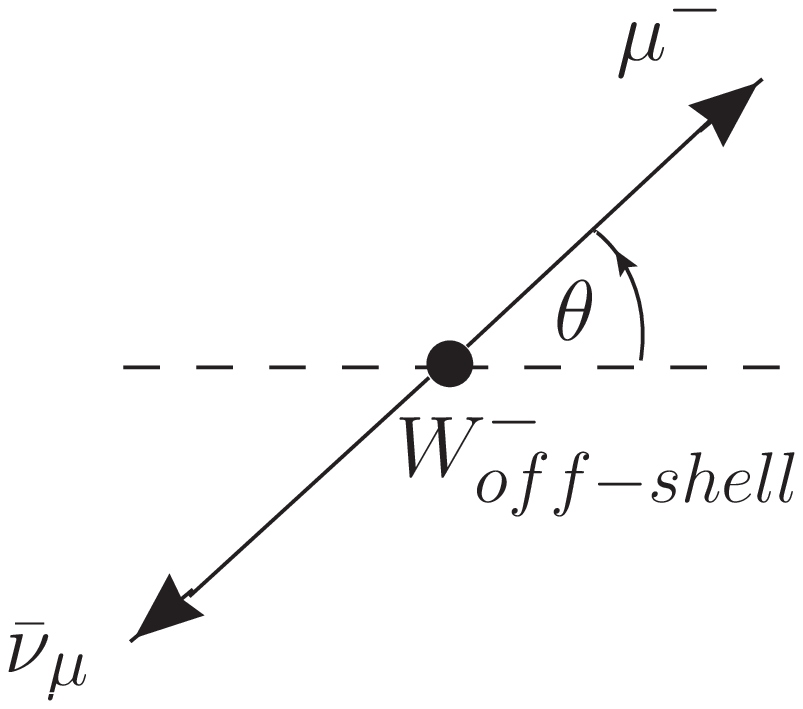, scale=0.30}
\end{center}
\end{figure}

\noindent
If $A_{FB}\neq 0$ one speaks of a parity-odd effect. Consider the $J^P$
content of the currents coupling to the $W^\ast$:\quad $V^\mu(1^-,0^+)$
and $A^\mu(1^+,0^-)$. There are two sources of parity-odd effects leading to
$A_{FB}\neq 0$ given by
\begin{enumerate}
\item parity-violating interaction from $V(1^-)A(1^+)$
  interference
\item parity-conserving interaction from $V(0^+)V(1^-)$, $A(0^-)A(1^+)$
  interference
\end{enumerate}
Take, for example, the semileptonic decay
$\Lambda_b\to\Lambda_c+\ell^-\bar\nu_\ell$. The numerator of Eq.~(\ref{fb}) is
given by (see Ref.~\cite{Kadeer:2005aq})
\begin{equation}\label{fb1}
\frac{d\Gamma_F}{dq^2}-\frac{d\Gamma_B}{dq^2}=\frac{G^2}{(2\pi)^3}|V_{bc}|^2
  \frac{(q^2-m_\ell^2)^2p}{8M_1^2q^2}\left[-H_{\frac121}^VH_{\frac121}^A
  -2\frac{m_\ell^2}{2q^2}(H_{\frac12t}^VH_{\frac120}^V
  +H_{\frac12t}^AH_{\frac120}^A)\right].
\end{equation}
The amplitudes $H^{V,A}_{\lambda_{\Lambda_c}\lambda_W}$ in Eq.~(\ref{fb1}) 
denote
the helicity amplitudes in the transitions $\Lambda_b(\lambda_{\Lambda_b})\to
\Lambda_c(\lambda_{\Lambda_c})+W^{-\ast}(\lambda_W)$. The first term in
Eq.~(\ref{fb1}) arises from a  truly parity-violating contribution while the
remaining two contributions are parity-odd contributions arising from parity
conserving interactions. The second contribution is negligible for the $e^-$
and $\mu^-$ modes due to the helicity flip factor $m_\ell^2/q^2$, but can be
sizeable for the $\tau^-$ mode. In fact, for the $\tau$ mode
$\Lambda_b\to\Lambda_c+\tau^-\bar\nu_\tau$ the FB asymmetry is dominated by
the helicity flip contribution in Eq.~(\ref{fb1}) leading to a sign change in
$A_{FB}$ when going from the $e^-,\mu^-$ modes to the $\tau^-$ mode (see the
corresponding quark-level calculation in Ref.~\cite{Korner:1989qb}).

\section{Helicity amplitudes and invariant amplitudes}
The results of a dynamical calculation are usually obtained in terms of
invariant amplitudes. The helicity amplitudes can be expressed as a linear
superposition of the invariant amplitudes. In this section we show how to
calculate the coefficient of this linear expansion for the process
$B_1\to B_2+j_{\rm eff}$. In order to calculate the coefficients of the linear
expansion one has to choose a definite frame.

\subsection{System 1: Parent baryon $B_1$ at rest}
Consider the decay $B_1(M_1)\to B_2(M_2)+j_{\rm eff}$ in the rest system of
$B_1$. The effective current $j_{\rm eff}$ with momentum $q^\mu$ moves in the
positive $z$ direction while $B_2$ moves in the negative $z$ direction.
\begin{figure}[!htb]
\centering
\includegraphics[width=50mm]{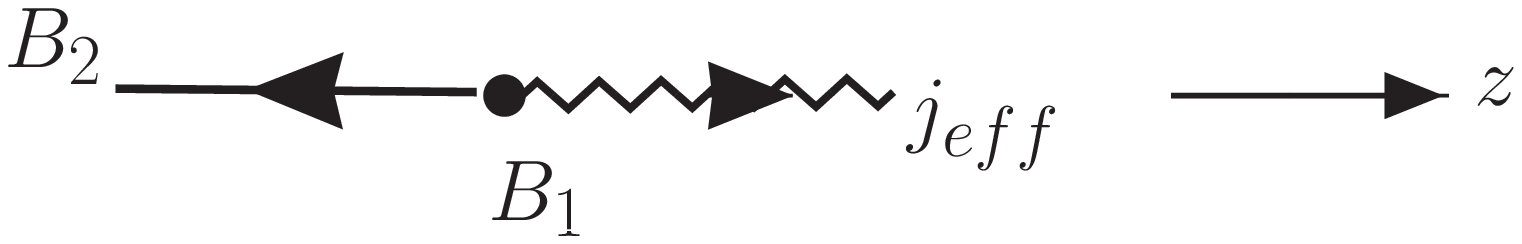}
\end{figure} 
\begin{equation}
p_1=(M_1;0,0,0)\qquad
q^\mu=(q_0;0,0,|\vec q\,|)\qquad
p_2^\mu=(E_2;0,0,-|\vec q\,|)\nonumber
\end{equation}
We do not explicitly annote the helicity of the parent baryon $B_1$ in the
helicity amplitudes since, in system 1, $\lambda_1$ is fixed by the relation 
$\lambda_1=-\lambda_2+\lambda_j$\,.

Possible helicity configurations are 
\begin{center}
\begin{tabular}{ccc}\hline\hline
$\lambda_{1}$ &$\lambda_{2}$ &$\lambda_{j}$ \\ \hline\hline
1/2 &-1/2 & 0\,\,(t) \\
-1/2 & 1/2 &0\,\,(t) \\
1/2 &1/2 &1 \\
-1/2 &-1/2&-1 \\ \hline
\end{tabular} 
\end{center}

\noindent
Convenient relations in system 1 are ($Q_\pm=(M_1\pm M_2)^2-q^2$)
\begin{equation}
2M_1(E_2+M_2)=Q_+,\qquad
2M_1|\vec q\,|=\sqrt{Q_+Q_-}.
\end{equation}
The helicity spinors are given by
\begin{eqnarray}\label{spinor1}
\bar u_2(\pm\frac12,p_2)=\sqrt{E_2+M_2}\left(\chi_\mp^\dagger,
  \frac{\mp|\vec q\,|\,}{E_2+M_2}\chi_\mp^\dagger\right),\quad
u_1(\pm\frac12,p_1)=\sqrt{2M_1}
  \left(\begin{array}{c}\chi_\pm\\0\end{array}\right),
\end{eqnarray}
where $\chi_+=\left(\begin{array}{c}1\\0\end{array}\right)$ and 
$\chi_-=\left(\begin{array}{c}0\\1\end{array}\right)$ are the usual Pauli 
two-spinors.

The helicity spinors satisfy the relations
\begin{eqnarray}
\frac12(1+\gamma_5s\!\!\!/_\pm)u(\pm\frac12,p)&=&u(\pm\frac12,p),\\
\frac12(1+\gamma_5s\!\!\!/_\mp)u(\pm\frac12,p)&=&0,\nonumber
\end{eqnarray}
where $s_{\mu\,\pm}=\pm(|\vec p\,|/M;0,0,E/M)$ is the spin four-vector of the
fermion with helicity $\pm1/2$.

For the four polarization four-vectors of the effective current we have 
\begin{equation}\label{polvec}
\varepsilon^\mu(t)=\frac1{\sqrt{q^2}}\left(q_0;0,0,|\vec q\,|\right),\quad
\varepsilon^\mu(\pm 1)=\frac1{\sqrt{2}}\left(0;\mp 1,-i,0\right),\quad
\varepsilon^\mu(0)=\frac1{\sqrt{q^2}}\left(|\vec q\,|;0,0,q_0\right). 
\end{equation}
They can be obtained by boosting the corresponding rest frame polarization
vectors $\varepsilon^\mu(t;q=0)=(1;0,0,0)$ and
$\varepsilon^\mu(0;q=0)=(0;0,0,1)$ by a boost with the nonvanishing elements
of the boost matrix given by $M_{tt}=M_{00}=q_0/\sqrt{q^2}$ and
$M_{t0}=M_{0t}=|\vec q\,|/\sqrt{q^2}$ (the transverse polarization vectors are
boost invariant).  

One defines helicity amplitudes through
\begin{equation}\label{heldef}
H^{V,A}_{\lambda_2\lambda_W} =M_\mu^{V,A}(\lambda_2)\epsilon^{*\mu}(\lambda_j).
\end{equation}
The current matrix elements can be expanded in terms of a complete set of
invariants
\begin{eqnarray}\label{expan}
M_\mu^V&=&\langle B_2|J_\mu^V|B_1\rangle
  = \bar u_2(p_2)\bigg[F_1^V(q^2)\gamma_\mu-\frac{F_2^V(q^2)}{M_1}
  i\sigma_{\mu\nu}q^\nu+\frac{F_3^V(q^2)}{M_1}q_\mu\bigg]u_1(p_1),\\
M_\mu^A&=&\langle B_2|J_\mu^A|B_1\rangle=\bar u_2(p_2)\bigg[F_1^A(q^2)
  \gamma_\mu-\frac{F_2^A(q^2)}{M_1}i\sigma_{\mu\nu}q^\nu
  +\frac{F_3^A(q^2)}{M_1}q_\mu\bigg]\gamma_5u_1(p_1)\nonumber
\end{eqnarray}
(we define
$\sigma_{\mu\nu}=\frac i2(\gamma_\mu\gamma_\nu-\gamma_\nu\gamma_\mu)$). Using
the definitions~(\ref{heldef},\,\ref{expan}), the helicity
spinors~(\ref{spinor1}) and polarization vectors~(\ref{polvec}), the helicity
amplitudes can be calculated to be
\begin{eqnarray}
\label{helinv}
H_{\frac12t}^{V/A}&=&\frac{\sqrt{Q_\pm}}{\sqrt{q^2}}
  \bigg((M_1\mp M_2)F_1^{V/A}\pm q^2/M_1F_3^{V/A}\bigg),\\
H_{\frac121}^{V/A}&=&\sqrt{2Q_\mp}
  \bigg(F_1^{V/A}\pm(M_1\pm M_2)/M_1F_2^{V/A}\bigg),\nonumber\\
H_{\frac120}^{V/A}&=&\frac{\sqrt{Q_\mp}}{\sqrt{q^2}}
  \bigg((M_1\pm M_2)F_1^{V/A}\pm q^2/M_1F_2^{V/A}\bigg).\nonumber 
\end{eqnarray}

From parity or from an explicit calculation one has
\begin{eqnarray*}
H_{-\lambda_2,-\lambda_j}^V&=&H_{\lambda_2,\lambda_j}^V,\\
H_{-\lambda_2,-\lambda_j}^A&=&-H_{\lambda_2,\lambda_j}^A.
\end{eqnarray*}
For a general linear combination $H_{\lambda_2,\lambda_j}
=aH_{\lambda_2,\lambda_j}^V+bH_{\lambda_2,\lambda_j}^A$ it is advantageous to
make use of the linear superpositions
$(H_{\lambda_2,\lambda_j}\pm H_{-\lambda_2,-\lambda_j})$ which have definite
transformation properties under parity. For example, it is convenient to
define so-called transversity amplitudes for the transverse helicities 
$\lambda_j=\pm1$ via $A_{\,\parallel,\perp}=(H_{\lambda_2,\lambda_j}
\pm H_{-\lambda_2,-\lambda_j})/\sqrt{2}$\,.

\subsection{System 2: The effective current is at rest}
The effective current $j_{\rm eff}$ is at rest, or put differently, in
system 2 we work in the cm frame of the lepton pair in the decay
$j_{\rm eff}\to \ell\bar\ell$. Both $B_1$ and $B_2$ move in the negative $z$
direction. One now has $\lambda_1=\lambda_2-\lambda_j$.
\begin{figure}[!htb]
\centering
\includegraphics[width=50mm]{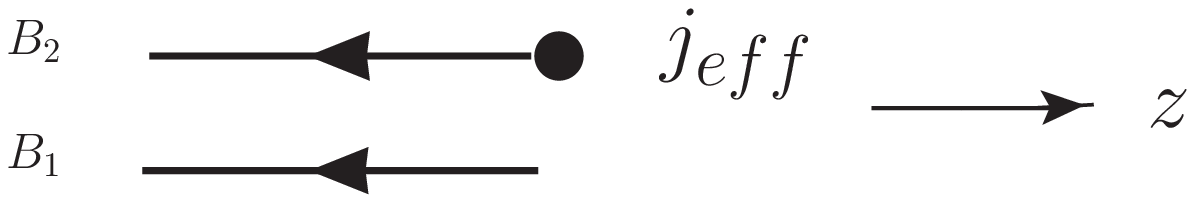}
\end{figure} 
\begin{equation}
p_1^\mu=(E'_1;0,0,-|\vec p\,'|)\qquad
q^\mu=(\sqrt{q^2};0,0,0)\qquad
p_2^\mu=(E'_2;0,0,-|\vec p\,'|)\nonumber
\end{equation}

Convenient relations in system 2 are
\begin{equation}
|\vec{p}\,'|=\frac{\sqrt{Q_+Q_-}}{2\sqrt{q^2}},\quad
(E'_1+M_1)(E'_2+M_2)=\frac{Q_+}{4q^2}\,(M_1-M_2+\sqrt{q^2})^2.
\end{equation}
The relevant spinors can be obtained from the rest frame spinor in
Eq.~(\ref{spinor1}) by a boost according to
$2M_1u(p_1)=(p\!\!\!/_1+M_1)u(p_1=0)$ and 
$2M_1\bar u(p_1)=\bar u(p_1=0)(p\!\!\!/_1+M_1)$. The spinors in system 2 are
thus given by
\begin{equation}
\bar u_2(\pm\frac12,p_2)=\sqrt{E'_2+M_2}\left(\chi_\mp^\dagger,
  \frac{\mp|\vec p\,'|}{E'_2+M_2}\chi_\mp^\dagger\right),\quad
u_1(\pm\frac12,p_1)=\sqrt{E'_1+M_1}\left(\begin{array}{c}\chi_\mp\\ 
  \frac{\pm|\vec p\,'|}{E'_1+M_1}\chi_\mp\end{array}\right)\nonumber 
\end{equation}
For the four polarization four-vectors of the effective current we now have
$\varepsilon^\mu(t)=\left(1;0,0,0\right)$ and
$\varepsilon^\mu(0)=\left(0;0,0,1\right)$ while the transverse polarization
vectors $\varepsilon^\mu(\pm 1)=\frac1{\sqrt2}\left(0;\mp 1,-i,0\right)$
remain unchanged.

With a little bit of work one can show that
\begin{equation}
H_{\lambda_2,\lambda_j}^{V,A}(\mbox{system 2})\,\,(F_i^{V,A})=
H_{\lambda_2,\lambda_j}^{V,A}(\mbox{system 1})\,\,(F_i^{V,A}),\nonumber
\end{equation}
i.e.\ Eq.~(\ref{helinv}) holds for both systems 1 and 2. One has recovered a
general property of the linear coefficients relating the helicity amplitudes
to invariant amplitudes: the coefficients of this linear relation are boost
invariant. In this sense the helicity amplitudes are boost invariant. I have
gone through this exercise in some detail to convince the reader that e.g.\
the expression $\sum_{\lambda_2}|H_{\lambda_2,\lambda_j}|^2$ is nothing but
the (unnormalized) density matrix of the off-shell gauge boson in its own rest 
frame regardless of the system in which the helicity amplitudes are evaluated
(as long as the systems are connected by a boost). We mention that
corresponding relations between helicity amplitudes and invariant amplitudes
for the cases $(1/2^+;3/2^+)\to(1/2^+;3/2^+)$ have been given in
Ref.~\cite{Faessler:2009xn}.

\subsection{Helicity amplitudes and $(LS)$ amplitudes}
Looking at Eq.~(\ref{helinv}) one notes that at threshold $q^2=(M_1-M_2)^2$
there are only two independent nonvanishing helicity amplitudes, namely
$H^V_{1/2,t}$ and $H^A_{1/2,1}=\sqrt 2H^A_{1/2,0}$. This is no accident and
can be understood by performing an $(LS)$ amplitude analysis in terms of the
$(LS)$ amplitudes $A^{V,A}_{LS}$. For the vector component with $J^P$ content
$(1^-;0^+)$ one has the $(LS)$ amplitudes
$(A^V_{1,1/2},A^V_{1,3/2};A^V_{0,1/2})$, and for the axial component with
$J^P$ content $(1^+;0^-)$ one has the $(LS)$ amplitudes 
$(A^A_{0,1/2},A^A_{2,3/2};A^A_{1,1/2})$. At threshold only the two $S$-wave
amplitudes survive, namely $A^V_{0,1/2}$ and $A^A_{0,1/2}$. In fact, there is
a linear relation between the set of helicity and $(LS)$ amplitudes which
reads ($J=0,1$)
\begin{equation}\label{lsampl}
H_{\lambda_1\lambda_2}(J)=\sum_{LS}\left(\frac{2L+1}{2J+1}\right)^{1/2}
  \langle LS0\mu|J\lambda\rangle\langle s_1s_2
  -\lambda_2\lambda_1|S\mu\rangle\,A_{LS},
\end{equation}
where $\lambda=\lambda_1-\lambda_2$. Eq.~(\ref{lsampl}) can be inverted, and
upon setting $A^A_{2,3/2}=0$ at threshold one recovers the above threshold
relation $H^A_{1/2,1}=\sqrt 2H^A_{1/2,0}$. We emphasize that the set of
$(LS)$ amplitudes is completely equivalent to the set of helicity amplitudes
and the definition of both sets of amplitudes is based on fully relativistic
concepts. Some examples of threshold and near threshold relations have
recently been discussed in
Refs.~\cite{Groote:2010nk,Zwicky:2013eda,Hiller:2013cza}. 

\section{Rotation of density matrices}
For concreteness we discus the decay of an on-shell $W^+$ into a fermion pair,
i.e.\ $W^+\to\bar f_3f_4$ (as e.g.\ $W^+\to\mu^+\nu_\mu$) described by the
helicity amplitudes $h_{\lambda_3\lambda_4}$ ($\lambda_3,\lambda_4=\pm1/2$).
First consider a frame where $W^+$ is at rest and where the antifermion 
$\bar f_3$ moves in the positive $z'$ direction.
\begin{itemize}
\item Consider first the decay of an unpolarized $W^+$ into a fermion pair. 
The decay rate in the $z'$ frame is given by
\begin{equation}
\Gamma\quad\sim\quad\sum_{\rm helicities}|h_{\lambda_3\lambda_4}|^2.\nonumber
\end{equation}
\item Consider next the decay of a polarized $W^+$ into a fermion pair. The
polarization of the $W^+$ is given in terms of the spin density matrix
$\rho'_{mm}$ with $m=\lambda_3-\lambda_4$. One then has 
\begin{equation}
\Gamma\quad\sim\quad\sum_{\rm helicities}\rho'_{mm}|h_{\lambda_3\lambda_4}|^2.
\nonumber
\end{equation}
\begin{figure}[!htb]\begin{center}
\epsfig{figure=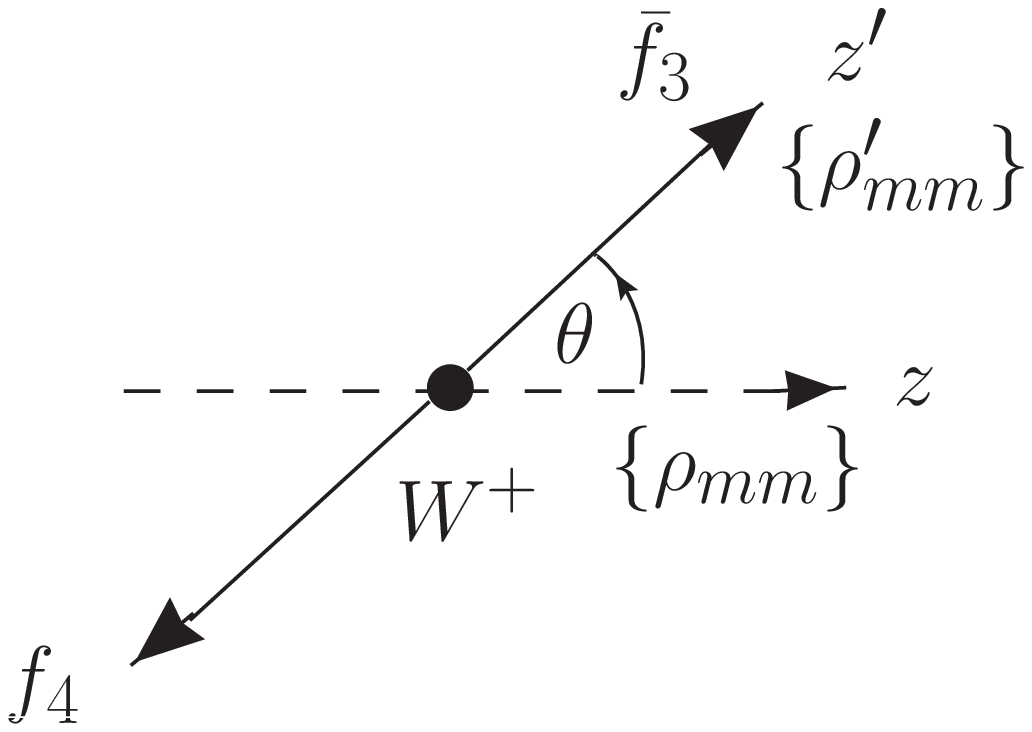, scale=0.35}
\end{center}\end{figure}
\item Now assume that the $W^+$ was polarized in a production process
characterized by a $z$ axis as e.g.\ in the decay $t\to b+W^+$ discussed
before. In this case the $z$ axis is defined by the momentum direction of the
$W^+$ in the top quark rest system. The spin density matrix of the $W^+$ is
given in terms of the helicity amplitudes for the decay $t\to b+W^+$, i.e.\ by
$\sum_{\lambda_2}H_{\lambda_2\lambda_W}$. In the present case (no azimuthal
correlations) one only needs the diagonal terms of the density matrix of the
$W^+$. For the unnormalized density matrix elements of the $W^+$ one has
\begin{equation}
\rho_{m=\lambda_W,\,m=\lambda_W}=\sum_{\lambda_2}|H_{\lambda_2\lambda_W}|^2.
\nonumber
\end{equation}   
Then ``rotate'' the density matrix. Rotation is from $(x,y,z)$ to $(x',y,z')$
by the angle $\theta$ around the $y$ axis. The differential $\cos\theta$ rate
reads
\begin{equation}\label{theta1}
\frac{d\Gamma(\theta)}{d\cos\theta}\,\sim \,
  \sum_{\rm helicities}|h_{\lambda_3\lambda_4}|^2
  \underbrace{d^1_{\lambda_W,\lambda_3-\lambda_4}(\theta)
  \rho_{\lambda_W,\lambda_W}d^1_{\lambda_W,\lambda_3-\lambda_4}
  (\theta)}_{\mbox{rotated density matrix $\rho'$}}
\end{equation}
\end{itemize}

\subsection{General polarized two-body decay}
\begin{itemize}
\item Take the two particle decay $a\to b+c$ of a spin-$J_a$ particle where
the polarization of particle $a$ in the frame $(x,y,z)$ is given by
$\rho_{\lambda_a\lambda'_a}$. Since we are also considering possible effects
from azimuthal correlations one has to take into account the nondiagonal
density matrix elements $\rho_{\lambda_a\lambda'_a}$ with
$\lambda_a\neq\lambda'_a$.
\item Consider a second frame $(x',y',z')$ obtained from $(x,y,z)$ by the
rotation $R(\theta,\phi,0)$ and whose $z$ axis is defined by particle $b$. The
polarization density matrix $\rho'$ in the frame $(x',y',z')$ is obtained by a
``rotation'' of the density matrix $\rho$ from the frame $(x,y,z)$ to the
frame $(x',y',z')$. 
\item The rate for $a\to b+c$ is then given by the sum of the decay
probabilities $|h_{\lambda_b\lambda_c}|^2$ (with
$\lambda_a=\lambda_b-\lambda_c$) weighted by the diagonal terms of the density
matrix $\rho'$ of particle $a$ in the frame $(x',y',z')$. One has
\begin{equation}\label{basic}
\frac{d\Gamma_{a\to b+c}}{d\cos\theta\,d\phi}\,\sim
  \sum_{\lambda_a,\lambda'_a,\lambda_b,\lambda_c}|h_{\lambda_b\lambda_c}|^2 
  \underbrace{D^{J*}_{\lambda_a,\lambda_b-\lambda_c}(\theta,\phi)\,\,
  \rho_{\lambda_a,\lambda'_a}\,D^J_{\lambda'_a,\lambda_b-\lambda_c}
  (\theta,\phi)}_{\mbox{rotated density matrix $\rho'$}}
\end{equation}
where
\begin{equation}
D^J_{m,m'}(\theta,\phi)=e^{-im\phi}d^J_{m\,m'}(\theta).\nonumber
\end{equation}
\item All master formulas discussed in this lecture can be obtained by a
repeated application of the basic two-body formula.
\end{itemize}

\section{T-odd contributions}
Take again the cascade decay
$\Xi^0\to\Sigma^+(\to p\pi^0)+W^{-\ast}(\to\ell^-\nu_\ell)$ as an example.
Using the master formula Eq.~(\ref{five1}) one obtains, among others,
contributions from the two helicity configurations~\cite{Kadeer:2005aq} 
\begin{equation}
(\lambda_\Sigma=1/2,\lambda_W=1;\lambda'_\Sigma=-1/2,\lambda'_W=0)
\quad\mbox{and}\quad
(\lambda_\Sigma=-1/2,\lambda_W=0;\lambda'_\Sigma=1/2,\lambda'_W=1).\nonumber
\end{equation}
These will lead to the bilinear combinations
\begin{eqnarray*}
\lefteqn{H_{\frac121}H^*_{-\frac120}\,e^{i(\pi-\chi)}
  +H_{-\frac120}H^*_{\frac121}\,e^{-i(\pi-\chi)}}\nonumber\\
  &=&-2\cos\chi\,{\rm Re\,}H_{\frac121}H^*_{-\frac120}
  -2\sin\chi\,{\rm Im\,}H_{\frac121}H^*_{-\frac120}\,.
\end{eqnarray*}
Take the imaginary part contributions and put in the remaining $\theta$- and
$\theta_B$-dependent trigonometric factors. One has two terms proportional to
$\sin\chi$,
\begin{equation}\label{im1}
\sin\theta\sin\chi\sin\theta_B\,{\rm Im\,}H_{\frac121}H^*_{-\frac120}
  \qquad\mbox{and}\qquad
\cos\theta\sin\theta\sin\chi\sin\theta_B\,
  {\rm Im\,}H_{\frac121}H^*_{-\frac120}.
\end{equation}
 
\begin{figure}[t]\begin{center}
\epsfig{figure=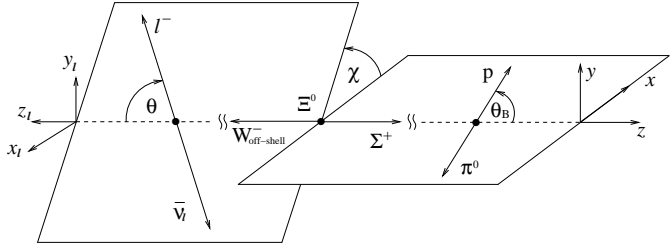, scale=0.40}
\caption{\label{five-fold} Definition of the polar angles $\theta$ and
  $\theta_B$, and the azimuthal angle $\chi$ in the joint angular decay
  distribution of an unpolarized $\Xi^0$ in the cascade decay
  $\Xi^0\to\Sigma^+(\to p+\pi^0)+\ell^-+\bar\nu_l\ell$. The coordinate system
  $(x_\ell,y_\ell,z_\ell)$ is obtained from the coordinate system $(x,y,z)$ by
  a $180^\circ$ rotation around the $y$ axis.}
\end{center}\end{figure}

Rewrite the product of angular factors in terms of scalar and pseudoscalar
products using the momentum representations in the $(x,y,z)$ system. The
normalized three-momenta are given by (see Fig.~\ref{five-fold})
\begin{eqnarray*}
\hat p_{\ell^-}&=&(\sin\theta\cos\chi,\sin\theta\sin\chi,-\cos\theta),\qquad
\hat p_W=(0,0,-1),\\
\hat p_{\Sigma^+}&=&(0,0,1),\qquad
\hat p_p=(\sin\theta_B,0,\cos\theta_B),
\end{eqnarray*}
where the three-momenta have unit length indicated by the hat notation. 

The two angular factors~(\ref{im1}) can be rewritten in terms of scalar
and cross products as
\begin{eqnarray}
\sin\theta\sin\chi\sin\theta_B &=&\hat p_W\cdot\left(\hat p_{\ell^-}
  \times\hat p_p\right),\label{t-odd1}\\
\cos\theta\sin\theta\sin\chi\sin\theta_B&=&(\hat p_{\ell^-}\cdot\hat p_W)\,  
  [\,\hat{p}_W\cdot\left(\,\hat p_{\ell^-}\!\times\hat p_p\right)]
  \label{t-odd2}
\end{eqnarray}  
Under time reversal ($t\to-t$) one has $(\hat p\to-\hat p)$. The above two
invariants~(\ref{t-odd1}) and~(\ref{t-odd2}) involve an odd number of momenta,
i.e.\ they change sign under time reversal. This has led to the notion of the
so-called $T$-odd observables: Observables that multiply $T$-odd momentum
invariants are called $T$-odd observables.

In the same vein we rewrite the angular factors multiplying $\cos\chi$. One
finds
\begin{eqnarray*}\label{t-odd}
\sin\theta\cos\chi\sin\theta_B&=&\hat p_W\cdot\hat p_p
  +\left(\hat p_W\cdot\hat p_p\right)\left(\hat p_W\cdot\hat p_{\ell^-}
  \right),\\
\cos\theta\sin\theta\cos\chi\sin\theta_B&=&(\hat p_{\ell^-}\cdot\hat p_W)\,
  \left(\hat p_W\cdot\hat p_p+\left(\hat p_W\cdot\hat p_p\right)
  \left(\hat p_W\cdot\hat p_{\ell^-}\right)\right).
\end{eqnarray*}
There is an even number of momentum factors in the angular correlations
involving $\cos\chi$, i.e.\ the momentum invariants correspond to $T$-even
angular correlations.

The $T$-odd contributions can arise from two different sources. They can be
contributed to by true $CP$-violating effects or by final state interaction
effects (imaginary parts of loop contributions). One can distinguish between
the two sources of $T$-odd effects by comparing with the corresponding
antihyperon decays. Phases from $CP$-violating effects change sign whereas
phases from final state interaction effects do not change sign when going from
hyperon to antihyperon decays.  

\section{Two examples of polar angle decay distributions}
\subsection{The top quark decay $t \to b +W^{+}(\to \ell^{+}+\nu_{\ell})$}
We are finally ready to derive the polar angle distribution
$W(\theta)\sim L_{\mu\nu}H^{\mu\nu} $ in the decay
$t\to b+W^+(\to\ell^++\nu_\ell)$ using helicity methods. The momentum 
dependent terms in Eq.~(\ref{topdecay}) can be dropped in the zero lepton
mass approximation. We take the $W^+$ to be on-shell, i.e.\ the $W^+$ has
three spin degrees of freedom with corresponding helicities
$\lambda_W=\pm1,0$. Heeding Eq.~(\ref{theta1}) one has
\begin{equation}
L_{\mu\nu}H^{\mu\nu}=\frac18\sum_{\lambda_b,\lambda_W,\lambda_\ell}
  |H^{V-A}_{\lambda_b\,\lambda_W}|^2\,
  d^1_{\lambda_W,\lambda_\ell+\frac12}(\theta)\,
  d^1_{\lambda_W,\lambda_\ell+\frac12}(\theta)\,
  |h^{V-A}_{\lambda_\ell,-\frac12}|^2. 
\end{equation}
At the scale of the process one can put the lepton-side helicity flip
amplitude to zero, i.e.\ $|h^{V-A}_{-\frac12,-\frac12}|^2=0$. The helicity
nonflip amplitude is given by $|h^{V-A}_{\frac12,-\frac12}|^2=8m_W^2$. One
obtains
\begin{equation}
L_{\mu\nu}H^{\mu\nu}=\frac{m_W^2}{4}\Big(|H^{V-A}_{\frac121}|^2
  (1+\cos\theta)^2+2(|H^{V-A}_{\frac120}|^2
  +|H^{V-A}_{-\frac120}|^2)\sin^2\theta
  +|H^{V-A}_{-\frac121}|^2(1-\cos\theta)^2\Big).\nonumber
\end{equation}
The corresponding three-fold angular decay distribution of polarized top decay
$t(\uparrow)\to b+W^+(\to\ell^++\nu_\ell)$~\cite{Fischer:1998gsa,%
Fischer:2001gp} can be derived with similar ease.

As emphasized in Sec.~3.2 the bilinear forms 
$\sum_{\lambda_b}|H^{V-A}_{\lambda_b\,\lambda_W}|^2$ 
($\lambda_j=1,0,-1$) are the (unnormalized) density matrix elements of the
on-shell $W^+$ in the $W^+$ rest frame. In their normalized form the density
matrix elements  $\sum_{\lambda_b}|\widehat H^{V-A}_{\lambda_b\,\lambda_j}|^2$
are usually referred to as the helicity fractions of the $W^+$ labelled by
${\cal H}_+$, ${\cal H}_0$ and ${\cal H}_-$. At the Born term level and for
$m_b=0$ one has ($y^2=m_W^2/m^2_t$)
\begin{equation}
{\cal H}_+:{\cal H}_0:{\cal H}_-=0:\frac1{1+2y^2}:\frac{2y^2}{1+2y^2}
  =0:0.70:0.30,
\end{equation}
where we have used $m_t=173.5$\,GeV. NLO and NNLO QCD corrections to the
helicity fractions have been calculated in
Refs.~\cite{Fischer:1998gsa,Fischer:2001gp} and in
Ref.~\cite{Czarnecki:2010gb}, respectively. Results on the NLO elctroweak
corrections to the helicity fractions have been given in Ref.~\cite{Do:2002ky}.

\subsection{The decay $\Lambda_b(\uparrow)\to\Lambda+J/\psi(\to\ell^+\ell^-)$}
\begin{figure}[t]
\begin{center}
\epsfig{figure=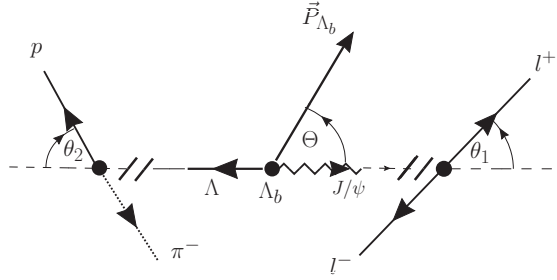, scale=0.35}
\caption{\label{three-fold} Definition of three polar angles in the decay
  $\Lambda_b(\uparrow)\to\Lambda(\to p+\pi^-)+J/\psi(\to\ell^+\ell^-)$}
\end{center}
\end{figure}

There has been a longstanding interest to measure the polarization of
hadronically produced hyperons, and charm and bottom
baryons~\cite{Lednicky:1985zx,Hrivnac:1994jx}. Recently the LHCb Collaboration
has measured the polarization of hadronically produced
$\Lambda_b$'s~\cite{Aaij:2013oxa}. At the same time they measured ratios of
squared helicity amplitudes in the decay $\Lambda_b(\uparrow)\to\Lambda+J/\psi$
through an analysis of polar correlations in the cascade decay process.
Consider the three polar angles $\theta$, $\theta_1$ and $\theta_2$ that
characterize the cascade decay
$\Lambda_b(\uparrow)\to\Lambda(\to p+\pi^-)+J/\psi(\to\ell^+\ell^-)$
(see Fig.~\ref{three-fold})

By now we know how to write down the master formula for this three-fold polar
angle distribution which could also be obtained by azimuthal integration of
Eq.~(\ref{five1}). Since I also want to discuss the decay
$\Lambda_b(\uparrow)\to\Lambda(\to p+\pi^-)+\psi(2S)(\to\ell^+\ell^-)$ I use
the generic notation $V$ for the $J^{PC}=1^{--}$ vector resonances $J/\psi$
and $\psi(2S)$. In the $\psi(2S)$ mode one also has access to the decay
$\psi(2S)\to\tau^+\tau^-$ which necessitates the incorporation of lepton mass
effects in the decay distribution. One has
\begin{eqnarray*}
W(\theta,\theta_1,\theta_2)
  &\propto&\frac12\sum_{helicities}|h^V_{\lambda_1\lambda_2}|^2
  \left[d^1_{\lambda_V,\lambda_1-\lambda_2}(\theta_2)\right]^2
  \rho_{\lambda_b,\lambda_b}(\theta)\\&&\times\
  \delta_{\lambda_b,\lambda_V-\lambda_\Lambda}
  |H_{\lambda_\Lambda\lambda_V}|^2\left[d^{1/2}_{\lambda_\Lambda\lambda_p}
  (\theta_1)\right]^2|h^B_{\lambda_p,0}|^2,
\end{eqnarray*}
where $\lambda_V$ is the helicity of the $J/\psi$ or $\psi(2S)$. The lepton
non-flip (n.f.) and flip (h.f.) helicity amplitudes for the parity conserving
decays $V\to\ell^+\ell^-$ are given by 
\begin{equation}
\mbox{n.f.}:\qquad h^V_{-\frac12,-\frac12}=h^V_{+\frac12,+\frac12}=2m_l,
  \qquad
\mbox{h.f.}:\qquad h^V_{-\frac12,+\frac12}=h^V_{+\frac12,-\frac12}=\sqrt2m_V.
\nonumber
\end{equation}

We also know how to rotate the density matrix of the $\Lambda_b$ from its
production direction (perpendicular to the production plane)
\begin{equation}\label{densitym}
\mathrm{\rho_{\lambda_b\lambda'_b}(\theta_P)}=\frac12\left(
  \begin{array}{cc}
    1+\mathrm{P}_b\cos{\theta_{\mathrm{P}}}
    &\mathrm{P}_b\sin{\theta_{\mathrm{P}}}\\
    \mathrm{P}_b\sin{\theta_{\mathrm{P}}}
    &1-\mathrm{P}_b\cos{\theta_{\mathrm{P}}}
  \end{array}\right)\nonumber
\end{equation}
Since I am not considering azimuthal correlations in this application only
the diagonal density matrix elements $\rho_{\lambda_b\lambda_b}$ are needed.

Next introduce the linear combinations of normalized helicity bispinors
$|\widehat H_{\lambda_\Lambda\lambda_V}|^2$ (where
$|\widehat H_{\lambda_\Lambda\lambda_V}|^2=|H_{\lambda_\Lambda\lambda_V}|^2/
\sum_{\lambda_\Lambda,\lambda_V}|H_{\lambda_\Lambda\lambda_V}|^2$)  
\begin{eqnarray*}
\alpha_b&=&|\widehat H_{+\frac120}|^2-|\widehat H_{-\frac120}|^2 
  +|\widehat H_{-\frac12-1}|^2-|\widehat H_{+\frac12+1}|^2\,,\\
r_0&=&|\widehat H_{+\frac120}|^2+|\widehat H_{-\frac120}|^2\,,\\
r_1&=&|\widehat H_{+\frac120}|^2-|\widehat H_{-\frac120}|^2\,.
\end{eqnarray*}
We define $\varepsilon=m_\ell^2/m_V^2$ such that the velocity of the lepton is
given by $v=(1-4\varepsilon)^{1/2}$.

The polar angle distribution can be written as
\begin{equation}\label{eq:w3}
\widetilde{W}(\theta,\theta_1,\theta_2)=\sum_{i=0}^7\!
  f_i(\alpha_b,r_0,r_1)\;g_i(P_b,\alpha_\Lambda)\;
  h_i(\cos\theta,\cos\theta_1,\cos\theta_2)\;\ell_i(\varepsilon). 
\end{equation}
The functions $f_i$, $g_i$, $h_i$ and $\ell_i$ are listed in the following
table. 

\begin{center}\begin{tabular}{llllc}
\hline\noalign{\vskip 2mm}
$i$ \hspace*{0.3cm} & $f_i(\alpha_b,r_0,r_1)$  \hspace*{0.3cm} & 
$g_i(P_b,\alpha_\Lambda)$  \hspace*{0.3cm} & 
    $h_i(\cos\theta,\cos\theta_1,\cos\theta_2)$ & $\ell_i(\varepsilon) $
\\ \noalign{\vskip 2mm}\hline\noalign{\vskip 2mm}
0   & $1$   & $1$    & $1$ &   $v\cdot(1+2\varepsilon)$
\\ \noalign{\vskip 1mm}
1   & $\alpha_b$  & $P_b$  & $\cos\theta$  &   
    $v\cdot(1+2\varepsilon)$
\\ \noalign{\vskip 1mm}
2   & $2 r_1-\alpha_b$  & $\alpha_\Lambda$  & $\cos\theta_1$  &   
      $ v\cdot(1+2\varepsilon)$
\\ \noalign{\vskip 1mm}
3   & $2 r_0-1$   & $ P_b\alpha_\Lambda$  & $\cos\theta\cos\theta_1$  & 
      $ v\cdot(1+2\varepsilon)$
\\ \noalign{\vskip 1mm}
4   & $\tfrac12(1-3 r_0)$ & $1$   & $\tfrac12(3\cos^2\theta_2-1)$   &   
      $v\,\cdot\,v^{2}$  
\\ \noalign{\vskip 1mm}
5   & $\tfrac12(\alpha_b-3 r_1)$  & $P_b$  &   
      $\tfrac12(3\cos^2\theta_2-1)\cos\theta$   &   
      $ v\,\cdot\,v^{2}$ 
\\ \noalign{\vskip 1mm}
6   & $-\tfrac12 (\alpha_b + r_1)$  & $\alpha_\Lambda$  &  
      $\tfrac12(3\cos^2\theta_2-1)\cos\theta_1$  &   
      $ v\,\cdot\, v^{2}$      
\\ \noalign{\vskip 1mm}
7   & $-\tfrac12 (1 + r_0)$    & $P_b\alpha_\Lambda$  &  
      $\tfrac12(3\cos^2\theta_2-1)\cos\theta\cos\theta_1$ &
      $ v\,\cdot\,v^{2}$ 
\\[1.5ex]
\hline\end{tabular}\end{center}

The symbols in the table are 
\begin{eqnarray*}
P_b\qquad &:&\mbox{polarization of $\Lambda_b$}\\
\alpha_b\qquad &:&\mbox{asymmetry parameter in the decay $\Lambda\to p+\pi^-$}
\end{eqnarray*}
The overall factor $v$ in the fifth column is the phase space factor for
$V\to\ell^+\ell^-$. The factors $(1+2\epsilon)$ ($S$-wave dominance) and
$v^2$ ($(S-D)$-wave interference) were calculated by us for the first time.
The LHCb Collaboration finds a very small polarization of the
$\Lambda_b$~\cite{Aaij:2013oxa}
\begin{equation}
P_{b}=0.05\pm0.07\pm0.02.\nonumber
\end{equation}
Our results on helicity amplitudes for the transitions
$\Lambda_b\to\Lambda$~\cite{Gutsche:2013oea} agree with the experimental
results~\cite{Aaij:2013oxa}. Our calculation is based on the confined
covariant quark model developed by us (see e.g.\ Refs.~\cite{Gutsche:2013pp,%
Gutsche:2013oea,Gutsche:2012ze,Ivanov:2011aa,Dubnicka:2010ev}).

\subsection{The confined covariant quark model in a nutshell}
The confined covariant quark model provides a field theoretic frame work for
the constituent quark model~(see e.g.\ Refs.~\cite{Gutsche:2013pp,%
Gutsche:2013oea,Gutsche:2012ze,Ivanov:2011aa,Dubnicka:2010ev}). Its main
features can be summarized as follows.

Particle transitions are calculated from Feynman diagrams involving quark
loops. For example, the $\Lambda_b\to\Lambda$ transition is described by a
two-loop diagram requiring a genuine two-loop calculation. The high energy
behaviour of quark loops is tempered by nonlocal Gaussian-type vertex
functions with a Gaussian-type fall-off behaviour. The particle-quark vertices
have interpolating current structure. Use free local quark propagators
$(m\,-\!\not\!p)^{-1}$ in the Feynman diagrams. The normalization of the
particle-quark vertices is provided by the compositeness condition which
embodies the correct charge normalization of the respective hadron. The
compositeness condition can be viewed as the field theoretic equivalent of the
normalization of the wave function of a quantum mechanical state. A universal
infrared cut-off provides for an effective confinement of quarks. There are
therefore no free quark poles in the Feynman diagrams.

HQET relations are recovered by using a static propagator for the heavy quark
($k_1$ is a loop momentum)
\begin{equation}
\frac1{m_b\,-\!\not\!k_1\,-\!\not\!p_1}
  \quad\to\quad\frac{1\,+\!\not\!v_1}{-2k_1v_1-2\bar\Lambda}.\nonumber
\end{equation}

\section{Summary}
The helicity method provides an easy and simple access to angular decay
distributions in sequential cascade decays. Polarization and mass effects are
readily incorporated. The corresponding techniques should belong to the basic
tool kit of every experimentalist and theorist working in particle physics
phenomenology.

\section*{Acknowledgements}
I would like to thank S.~Groote, T.~Gutsche, M.A.~Ivanov, V.E.~Lyubovitskij
and P.~Santorelli for a fruitful collaboration. I gratefully acknowledge the
support of S.~Groote and M.A.~Ivanov by the MITP Mainz while they were
visiting the University of Mainz. Thanks to B.~J\"ager for a clarifying
discussion on the use of the unitary gauge for an off-shell gauge boson. 


\end{document}